# The New Finite Temperature Schrödinger Equation

Li Heling[1※], Yang Bin[1,2], Xiong Ying[1]

1. School of Physics Electronics Information Engineering, Ningxia University, Yinchuan, 750021, China
2. Department of physics, Beijing Normal University, Beijing, 100056, China

※Corresponding author：Heling Li.

E-mail address：nigxiayclhl@163.com.

**Abstract:** Implant the thoughtway of thermostatistics in quantum mechanics, set up the new finite temperature Schrödinger equation, define the pure-state free energy, and revise the microscopic entropy introduced by Wu, *et al*.

## 1. Introduction

In 1974, S. Weinberg, *et al*. had shown how finite-temperature effects in a renormalizable quantum field theory could restore a symmetry which is broken at zero temperature [1, 2]. This marks the initiation of the finite-temperature field theory. In 1989, I. Kapusta and P. V. Landshoff systematized the finite-temperature field theory, and established the method that contacts the field theory with thermostatistics [3]. Recently, Xiangyao Wu, et al. defined the microscopic entropy based on the first law of thermodynamics, and tried to introduce the finite temperature Schrödinger equation [4]. We think the thought is a highlight to introduce finite temperature Schrödinger equation, but there is some inappropriate place in the introduction process of logical reasoning, so the equation introduced is not the best. Furthermore, the thoughtways of the thermostatistics in developing gravitation theory have shown some highlights [5, 6]. For these reasons, in this paper we attempt to introduce more appropriate new finite temperature Schrödinger equation based on the thermostatistics, define the pure-state free energy, and introduce new microscopic entropy.

## 2. The entropy in thermodynamic system

Introducing temperature in pure mechanics (Newtonian mechanics, quantum mechanics) is a thorny issue, because the temperature is a physical quantity that usually introduced and used in statistical thermodynamics（thermostatistics）, and is usually considered to be the measure of the thermal motion intensity of microscopic particles. The concept of temperature only exist in the problem related to "numerous" and "statistics", so the traditional pure mechanics generally does not involve temperature and is considered to be the theory about the motion law of objects in the absolute zero temperature. For example, Schrödinger equation is considered to be zero-temperature equation. While particles or objects generally move in the environment of a certain temperature, so we should consider the effects of the finite temperature. To introduce temperature in the Schrödinger equation, some of the concepts and methods of thermostatistics should be used.

Entropy is also a physical quantity comes from thermostatistics and usually appears matching with temperature, and the statistical temperature is defined as the partial derivative of entropy to internal energy. Moreover, the so-called microscopic entropy is first introduced in the introduction process of finite temperature Schrödinger equation, so it is necessary to recall the definition of entropy.

In quantum statistics, the entropy $S$ of a thermodynamic system with $N$ particles is defined by

$$S = <\hat{S}> = <-k_B \ln \hat{\rho}> = -k_B \text{Tr}(\hat{\rho} \ln \hat{\rho}) \tag{1}$$

where $\hat{\rho} = \sum_{i}^{W} |\psi_i\rangle P_i \langle\psi_i|$ is statistical operator, $k_B$ is Boltzmann constant, and $\hat{S} = -k_B \ln \hat{\rho}$ is entropy operator. $|\psi_i\rangle$ ($i=1,2,3…W$) is the wave function describing the $i$-th state of the system, where $W$ is the total number of the possible microscopic states, $P_i$ is the probability of the thermodynamic system in state $|\psi_i\rangle$. That is, the entropy is the statistical average value of the entropy operator. For a micro canonical system, $P_i=1/W$, $\hat{\rho}=1/W$. When the system is in the $j$-th pure state, $P_j=1$, $P_i=0$ ($i \neq j$), $\hat{\rho} = |\psi_j\rangle\langle\psi_j|$, $S=0$.

In classical statistics, the entropy is defined by the Boltzmann-Shannon

$$S = -k_B \sum_{i}^{W} P_i \ln P_i \tag{2}$$

For micro canonical system, as $P_i=1/W$, we have $S = k_B \ln W$. This is the famous Boltzmann relation.

## 3. Finite temperature Schrödinger equation

In thermostatistics, there is the following relationship between internal energy and entropy

$$U - TS = F \tag{3}$$

where $U$ is internal energy, $T$ is absolute temperature, and $F$ is free energy. In thermostatistics, the internal energy is defined as the statistical average value of mechanical energy (the sum of kinetic and potential energy) of the system. The free energy is named because it is the part of energy that can convert into work in internal energy.

For the point particle system with $N$ particles, the total mechanical energy of the system in the $i$-th state is

$$E_i = \sum_l^N \left(T_l^i + V_l^i\right) + \sum_{l<m}^N V_{lm}^i \tag{4}$$

where $T_l^i$ and $V_l^i$ are respectively the kinetic energy and potential energy of the $l$-th particle, and $V_{lm}^i$ is the interaction energy between the $l$-th and $m$-th particle of thermodynamical system in the $i$-th state. Substituting (2) and (4) into (3), we define $F = \sum_i^W P_i F_i$, where $F_i$ is the free energy of the multi-particle system in the $i$-th state, then we have

$$\sum_i^W P_i \left( \sum_l^N \left(T_l^i + V_l^i\right) + \sum_{l<m}^N V_{lm}^i \right) - T \left( -k_B \sum_i^W P_i \ln P \right)_i = \sum_i^W P_i F_i \tag{5}$$

therefore,

$$\sum_l^N \left(T_l^i + V_l^i\right) + \sum_{l<m}^N V_{lm}^i + k_B T \ln P_i = F_i, \ (i=1,2,3\ldots W) \tag{6}$$

The equation (6) is similar (but completely different) to the equation (15) in reference [3], but we think the equation (15) itself and its deviation in reference [3] is inappropriate (see Appendix).

When the system is in pure state $j$, $P_j=1$, $P_i=0$ ($i \neq j$), now we have

$$\sum_l^N \left(T_l^j + V_l^j\right) + \sum_{l<m}^N V_{lm}^j = F_j = E_j$$

for the states that $i \neq j$, the equation (6) is significant in the condition of absolute temperature $T=0$, we have

$$\sum_l^N \left(T_l^i + V_l^i\right) + \sum_{l<m}^N V_{lm}^i = F_i = E_i, \ (i \neq j)$$

Furthermore, when $T \to 0$, we have

$$\sum_l^N \left(T_l^i + V_l^i\right) + \sum_{l<m}^N V_{lm}^i = F_i = E_i, \ (i=1,2,3\ldots W) \tag{7}$$

This is precisely the expression of mechanical energy in Newtonian mechanics, when $T\to 0$ or in pure state, the free energy $F_i$ of the mechanical system in the $i$-th state would degenerate to the mechanical energy $E_i$ of the system. This not only explains why Newtonian mechanics is the theory about the motion law of objects in the absolute zero temperature, but also demonstrates that $F_i$ has the characteristics of mechanical energy at finite temperature. Therefore, when the mechanical quantities in (6) correspond to the operators in quantum mechanics, we have

$$F_i \to i\hbar \frac{\partial}{\partial t}, \quad T_l^i \to -\frac{\hbar^2}{2m_l}\nabla_l^2 \tag{8}$$

Noticing the $P_i$ is the probability (the probability of mixed ensemble, not of pure ensemble) of the thermodynamic system in state $|\psi_i\rangle$, and the original starting point of the statistical thermodynamics is the equal-probability assumption of isolated system. So we can choose $P_i=1/W$.

We can obtain

$$i\hbar\frac{\partial}{\partial t}|\psi_i\rangle = \left[\sum_l^N\left(-\frac{\hbar^2}{2m_l}\nabla_l^2 + V_l^i\right) + \sum_{l<m}^N V_{lm}^i - k_B T \ln W\right]|\psi_i\rangle \tag{9}$$

where $|\psi_i\rangle$ is the wave function of multi-particle system, and in the coordinate representation $|\psi_i\rangle = \psi(\vec{r}_1,\vec{r}_2,\cdots,\vec{r}_N,t)$. The equation (9) is finite temperature Schrödinger equation of multi-particle. When $T=0$, the equation (9) becomes Schrodinger equation of multi-particle system.

The partial differential (9) can be solved by the method of separation of variable. By setting

$$|\psi_i\rangle = \psi(\vec{r}_1,\vec{r}_2,\cdots,\vec{r}_N)g(t) \tag{10}$$

substituting (10) into (9), we have

$$\frac{dg(t)}{dt} = \frac{Eg(t)}{i\hbar}. \tag{11}$$

So

$$g(t) \propto \exp\left(\frac{-iEt}{\hbar}\right)$$

and

$$\left[\sum_l^N\left(-\frac{\hbar^2}{2m_l}\nabla_l^2 + V_l^i\right) + \sum_{l<m}^N V_{lm}^i - k_B T \ln W\right]\psi(\vec{r}_1,\vec{r}_2,\cdots,\vec{r}_N) = E\psi(\vec{r}_1,\vec{r}_2,\cdots,\vec{r}_N) \tag{12}$$

Equation (12) is time-independent finite temperature Schrödinger equation of multi-particle system.

When selecting $P_i$ to correspond to the probability of pure state $\langle\psi_i|\psi_i\rangle = |\psi_i|^2$ (we think this correspondence is too par-fetched), that is $P_i \to |\psi_i|^2$, we can obtain the finite temperature Schrödinger equation of multi-particle in reference [3] from the equation (6)

$$i\hbar \frac{\partial}{\partial t}|\psi_i\rangle = \left[\sum_l^N \left(-\frac{\hbar^2}{2m_l}\nabla_l^2 + V_l^i\right) + \sum_{l<m}^N V_{lm}^i + k_B T \ln|\psi_i|^2\right]|\psi_i\rangle \quad (13)$$

The equation (13) is obviously less appropriate and less concise than (9).

For the single-particle system, we define the single-particle free energy $f_i$ from the (6)

$$f_i = T^i + V^i + k_B T \ln p_i, \quad (i=1,2,3\ldots w) \quad (14)$$

where $T^i$ and $V^i$ are respectively the kinetic energy and potential energy of the single-particle in state $i$, and $p_i$ is the probability of the single-particle in state $i$. When this particle is in the isolated condition of thermodynamic systems, $p_i=1/w$, where $w$ is the micro-state (quantum state) number of a particle in isolated system. Corresponding the mechanical quantities in (14) to the operators in quantum mechanics

$$f_i \to i\hbar\frac{\partial}{\partial t}, \quad T^i \to -\frac{\hbar^2}{2m}\nabla^2 \quad (15)$$

we obtain the finite temperature Schrödinger equation of single-particle

$$i\hbar \frac{\partial}{\partial t}|\psi_i\rangle = \left(-\frac{\hbar^2}{2m}\nabla^2 + V^i - k_B T \ln W\right)|\psi_i\rangle \quad (16)$$

where $|\psi_i\rangle$ is the wave function of single-particle system, and we have $|\psi_i\rangle = \psi(\vec{r},t)$ in coordinate representation.

The partial differential (16) can be solved by the method of separation of variable. By setting

$$|\psi_i\rangle = \psi(\vec{r})g(t) \quad (17)$$

substituting (17) into (16), we have

$$\frac{dg(t)}{dt} = \frac{Eg(t)}{i\hbar}$$
$$g(t) \propto \exp\left(\frac{-iEt}{\hbar}\right) \quad (18)$$

and

$$-\frac{\hbar^2}{2m_l}\nabla^2\psi(\vec{r}) + V(r)\psi(\vec{r}) - (k_B T \ln W)\psi(\vec{r}) = E\psi(\vec{r}) \quad (19)$$

The equation (19) is time-independent finite temperature Schrödinger equation of single-particle system.

## 4. Microscopic entropy

If we need to introduce the microscopic entropy, noticing the equation (1) or (2), and the wave function is a probability wave, we could correspond $P_i$ to the pure-state probability $\langle \psi_i | \psi_i \rangle = |\psi_i|^2$, that is $P_i \to |\psi_i|^2$. In coordinate representation, both $|\psi(\vec{r}_1, \vec{r}_2, \cdots, \vec{r}_N)|^2$ and $|\psi(\vec{r}, t)|^2$ are probability density, so we have:

the single particle's microscopic entropy

$$S = -k_B \int |\psi(\vec{r}, t)|^2 \ln |\psi(\vec{r}, t)|^2 \, d^3\vec{r} \tag{20}$$

the multi-particle's microscopic entropy

$$S = -k_B \int |\psi(\vec{r}_1, \vec{r}_2, \cdots, \vec{r}_N)|^2 \ln |\psi(\vec{r}_1, \vec{r}_2, \cdots, \vec{r}_N)|^2 \, d^3\vec{r}_1 d^3\vec{r}_2 \cdots d^3\vec{r}_N \tag{21}$$

## 5. Conclusion and discussion

In this paper, we defined pure-state free energy of single particle and multi-particle in eqs.(14,6), and introduced the more appropriate and more concise finite temperature Schrödinger equations of single particle and multi-particle in eqs.(9,11,12,16,18,19) and microscopic entropy in eqs. (20, 21). These results are the revisions to finite temperature Schrödinger equations and microscopic entropy introduced by Xiang-Yao Wu, *et al.*.

In the derivation process of the finite temperature Schrödinger equations, we selected the "equal-probability assumption", that is $P_i=1/W$. This is just a relatively simple one in the variety of options, the readers who are interested in it can try other options.

## Appendix

We think there are some inappropriate places in reference [4] as follows:

(1) In reference [4], the author believes "in classical statistics, the entropy is defined by the Boltzmann $S = k_B \ln W = -k_B \ln \rho$", this is incorrect. In classical statistics, the right definition of entropy is shown in eq. (6) in this paper, that is $S = -k_B \sum_i^W P_i \ln P_i$.

(2) The error of entropy expression will inevitably lead to the inappropriate definition of "microscopic entropy" in reference [4] (see the eqs.(6~8) in reference [4]).

(3) The deviation of the main equation (15) in reference [4],

$\sum_{i}^{N}(T_i+V_i)+\sum_{i<j}^{N}V_{ij}-TS=E$, is inappropriate, shown as follows:

*a*. It is no need to let *dW*=0;

*b*. The equation (11), $U=\sum_{i}^{N}(T_i+V_i)+\sum_{i<j}^{N}V_{ij}$, the author called internal energy, is actually mechanical energy. Its statistical average value is internal energy.

*c*. The equation (14) in reference [4], $d\left(\sum_{i}^{N}(T_i+V_i)+\sum_{i<j}^{N}V_{ij}-T\sum_{i}^{N}S_i\right)=0$, is derived under the no work, isothermal conditions based on the first and second thermodynamic laws and assumed the sum of the former two items is internal energy. And then the equation (15) is obtained from that. But eq. (15) means the system can get or loss any number of mechanical energy (meant internal energy by the author) only by heat transfer, while the temperature of the system is not change. This is impossible.

**Acknowledgments**

The financial support of the natural science foundation of Ningxia (NZ1023) is gratefully acknowledged.